# Implementation of Private Cloud using Eucalyptus and an open source Operating System

Nandan Mirajkar[1], Mohan Barde[2], Harshal Kamble[3], Dr.Rahul Athale[4], Kumud Singh[5]

[1]Department of Advanced Software and Computing Technologies
IGNOU – I²IT Centre of Excellence for Advanced Education and Research
Pune, Maharashtra 411 057, India

[2]Department of Advanced Software and Computing Technologies
IGNOU – I²IT Centre of Excellence for Advanced Education and Research
Pune, Maharashtra 411 057, India

[3]Department of Advanced Software and Computing Technologies
IGNOU – I²IT Centre of Excellence for Advanced Education and Research
Pune, Maharashtra 411 057, India

[4]Sunflower Information Technologies Pvt. Ltd Pune, Maharashtra 411 016, India

[5]Systems Department
IGNOU – I²IT Centre of Excellence for Advanced Education and Research
Pune, Maharashtra 411 057, India

**Abstract**
Cloud computing is bringing a revolution in computing environment replacing traditional software installations, licensing issues into complete on-demand services through internet. Microsoft office 365 a cloud based office application is available to clients online hence no need to buy and install the software. On Facebook a social networking website, users upload videos which uses cloud provider's storage service so less hardware cost for clients.Virtualization technology has great contribution in advent of cloud computing. Paper describes implementation of Private Cloud using open source operating system Ubuntu 10.04 server edition, installation of Ubuntu Enterprise Cloud with Eucalyptus 1.6.2 and providing CentOS 5.3 operating system through cloud.

## 1. Introduction

Cloud computing is a model for enabling ubiquitous (present everywhere), convenient, on-demand network access to a shared pool of configurable computing resources e.g., networks, servers, storage, applications, and services that can be rapidly provisioned and released with minimal management effort or service provider interaction [1]. Types of Cloud computing models are Public cloud, Private cloud and Hybrid cloud. In Public cloud the resources are provided over internet to all clients. In Private cloud resources are provided over intranet within an organization. Hybrid cloud is a provision depending on requirement can provide resources within an organization or publicly [7].Types of services in cloud computing include IaaS, SaaS and PaaS. In IaaS i.e Infrastructure as a service the virtual machines, raw (block) storage, firewalls, load balancers, and networks are offered which will navigate to provide cloud based services to clients. PaaS Platform as service is a way to provide platforms like operating systems, application development platforms e.g Microsoft's Visual Studio .Net to clients through internet. Software as Service SaaS enables service provider to give softwares like Enterprise Resource Planning (ERP), Customer Relationship Management (CRM) to users. CRM is still the largest market for SaaS with revenue to reach as predicted $3.8b in 2011, up from $3.2b in 2010 [6]

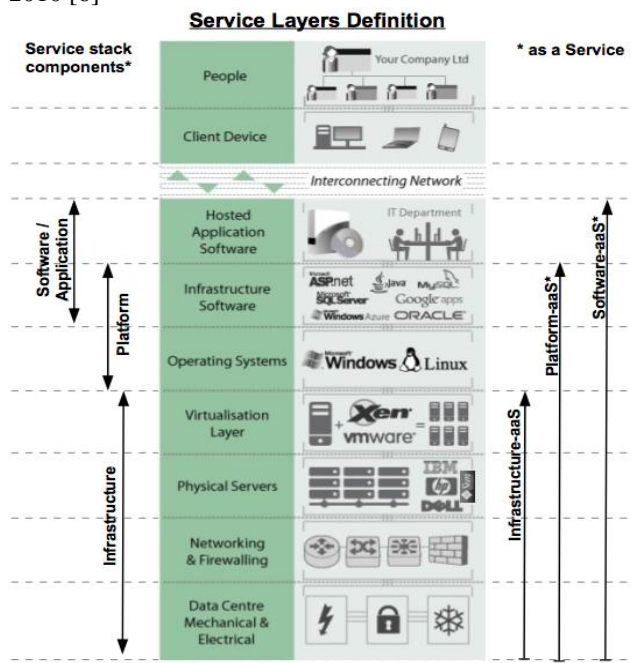

Fig 1: Services in Cloud[11]

## 2. Virtualization

Infrastructure as Service IaaS is base in cloud environment which provides Virtualization setup to create multiple workstations. In computing, virtualization means to create a virtual version of a device or resource, such as a server, storage device, network or even an operating system where the framework divides the resource into one or more execution environments [2] Hypervisor, also called virtual machine manager (VMM), is one of many hardware virtualization techniques allowing multiple operating systems, termed guests, to run concurrently on a host computer [15]. Bare-metal hypervisor is a thin layer of software that provides virtual partitioning methods, runs directly on hardware E.g are Citrix XenServer, VMware ESX/ESXi. Hosted is a virtualization technique where partitioning and virtualization services runs on top of a host operating system[3] E.g KVM-Kernel based Virtual Machine,VMware Player and VirtualBox.

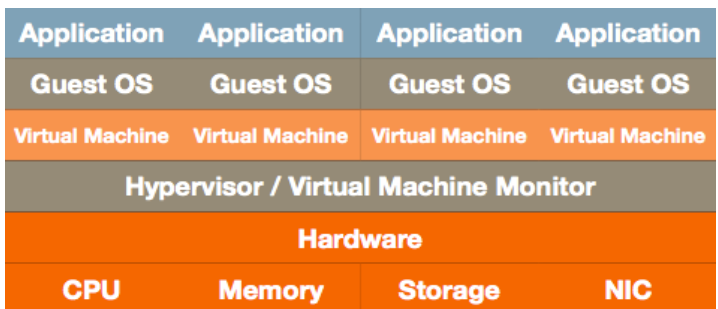
Fig 2: Bare-Metal Hardware Virtualization.[10]

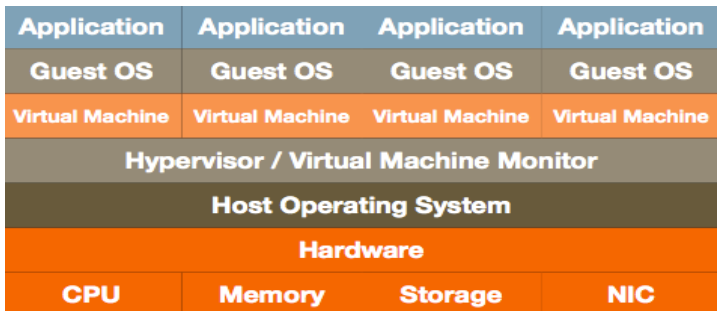
Fig 3: Hosted Virtualization.[10]

Full virtualization provides a complete simulation of underlying computer hardware, enabling software to run without any modification. Because it helps maximize the use and flexibility of computing resources, multiple operating systems can run simultaneously on the same hardware, full virtualization is considered a key technology for cloud computing. For cloud computing systems, full virtualization can increase operational efficiency because it can optimize computer workloads and adjust the number of servers in use to match demand, thereby conserving energy and information technology resources [4]. For Full virtualization emulation packages like VMware Server & Virtual Box are used. Para virtualization is a method for the hypervisor to offer interfaces to the guest operating system that the guest operating system can use, instead of the normal hardware interfaces. If a guest operating system can use paravirtualized interfaces, they offer significantly faster access for resources such as hard drives and networks [14]. Examples are Microsoft Hyper-V and VMware ESX Server.

## 3. Eucalyptus

Eucalyptus stands for 'Elastic Utility Computing Architecture for Linking Your programs to Useful Systems'. Eucalyptus is a research project under direction of Prof. Rich Wolski, Computer Science Department at the University of California, Santa Barbara, USA [6]. Now Eucalyptus is open source software used to create and manage a private or public cloud. Similar to Amazon's Elastic compute cloud EC2 Eucalyptus provides a cloud controller. Storage is Walrus which is similar to Amazon's S3 Simple Storage Service. Storage controller of Eucalyptus incorporates functionality of Elastic Block Storage EBS of Amazon. Cloud controller allows users and administrators to enter into cloud, for availability of resources it deals with node managers and acts as scheduler. Cluster controller schedules run-instance requests to particular node controllers and manage virtual instance network. Walrus is used to stream data in and out of cloud also from instances started on nodes. Virtual machine images can be stored and accessed using walrus. Storage controller manages dynamic block storage services. Node controller manages Virtual machine instances with the help of hypervisor [7].

## 4. Private Cloud using Ubuntu 10.04 server Edition

Hardware and software utilized:
1. Two Dell Rack servers R715 having same configuration for cloud and node controller as follows.
2. Two 1.8 GHz AMD processor:8 cores per processor
3. Internal storage- 300 GB
4. RAM - 8 GB
5. Virtualization support
6. Operating System – Ubuntu 10.04 server
7. Hypervisor- KVM
8. Cloud software- Eucalyptus Cloud 1.6.2

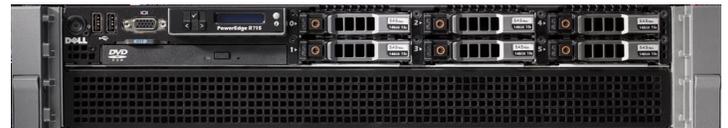
Fig 4: Dell Rack server R715 [12]

4.1 Setup server 1

1. Insert ubuntu 10.04 server edition cd
2. select 'Install ubuntu enterprise cloud'
3. Configure the network: select 'configure network manually'
address 192.168.0.221
gateway 192.168.0.1
netmask 255.255.255.0
nameserver 121.242.xxx.xxx
4. Host name for this system: cc
5. Cloud controller address : leave it blank
6. Cloud installation mode:
Select following Cloud controller, Walrus storage service, Cluster controller, Storage controller
7. Partition disks select 'Guided-use entries disk and set up LVM'
8. Full name for new user and username for account : cladmin
9. Select no automatic updates
10. Eucalyptus cluster name : cluster1
11. Pool of IP addresses that can be dynamically assigned as public IP's of virtual machines: 192.168.0.70-192.168.0.80

12. Install grub boot loader to master boot loader: yes[8]
Also install KVM on server1 which helps to install images and bundle them.
$ sudo apt-get install qemu-kvm [9]

4.2 Setup server 2

1. Insert ubuntu 10.04 server edition cd
2. select 'Install ubuntu enterprise cloud'
3. Configure the network: select 'configure network manually'
address 192.168.0.222
gateway 192.168.0.221 (IP of cloud controller)
netmask 255.255.255.0
nameserver 121.242.xxx.xxx
Here Cloud controller is detected automatically
4. Host name for this system: nc
5. Cloud installation mode: Node controller
6. Partition disks select 'Guided-use entire disk and set up LVM'
7. Full name for new user and username for account : cladmin
8. select no automatic updates
9. Install grub boot loader to master boot loader: yes[8]

4.3 Exchange of Public SSH Keys

On node controller set a temporary password
$ sudo passwd eucalyptus
On cloud controller
$ sudo -u eucalyptus ssh-copy-id -i /var/lib/eucalyptus/.ssh/id_rsa.pub eucalyptus@192.168.0.222
On node controller remove temporary password
$ sudo passwd -d eucalyptus

4.4 Get credentials

On cloud controller: install credentials which consist of x.509 certificates and environment variables.
Open web browser and enter following url:
https://192.168.0.221:8443/#login
and click on credentials tab

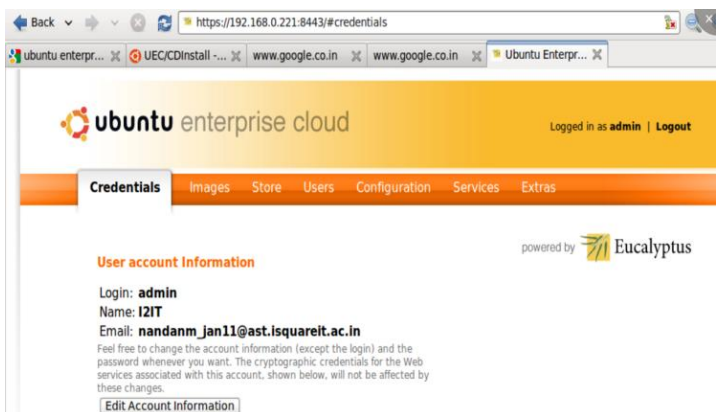
Fig 5: Screenshot of credentials tab

Download the credentials euca2-admin-x509.zip on cloud controller.
1.Copy credentials euca2-admin-x509.zip to /home/cladmin folder on the cloud controller.
$ scp euca2-admin-x509.zip /home/cladmin/
2. Create a hidden folder euca on the server1: cloud controller and extract the zip file here.
$ mkdir ~/.euca
$ cd ~/.euca
$ unzip /home/cladmin/euca2-admin-x509.zip
3. Remove the zip file for security reasons and apply permissions to the .euca folder and its contents.
$ rm ~/euca2-admin-x509.zip
$ chmod 0700 ~/.euca
$ chmod 0600 ~/.euca/*
4. Add following line to the ~/.bashrc file on the cloud controller so that necessary environment
variables are initialized upon login.
$ echo . ~/.euca/eucarc >> ~/.bashrc
5. Now source the .bashrc file to make sure settings take effect.
$ source ~/.bashrc [8]
9. Now install euca2ools if not installed, to communicate with the UEC
$ sudo apt-get install euca2ools
10. To check euca2ools is working try to get cluster1 availability details.
$ euca-describe-availability-zones verbose[8][13]

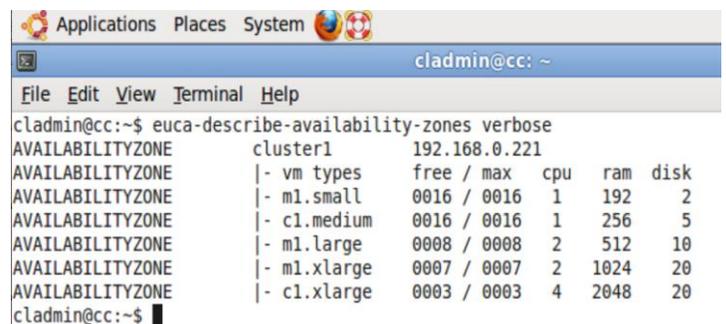
Fig 6: Screenshot for resources available

11. If the free/max VCPUs are 0000/0000 in figure 6 list it means node did not detect cloud controller during installation and in cloud controller, node is not registered automatically. On server 1: cloud controller using following command add 192.168.0.222 as node controller:
$ sudo euca_conf --discover-nodes [5]

4.5 Installing Images in Server1

The Store tab provides list of images, also it can be installed using following commands.

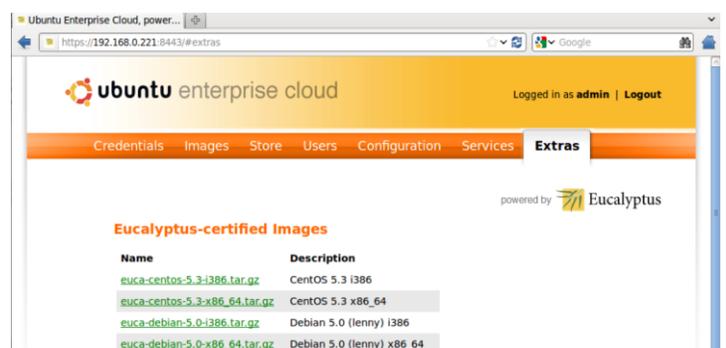
Fig 7: Images

Download images from Extras tab in UEC
1. Registering kernel image Execute the following commands to bundle and register the kernel image (vmlinuz-2.6.28-11-server):
$ euca-bundle-image -i vmlinuz-2.6.28-11-server --kernel true
$ euca-upload-bundle -b mybucket -m /tmp/vmlinuz-2.6.28-11-server.manifest.xml
$ euca-register mybucket/vmlinuz-2.6.28-11-server.manifest.xml
IMAGE eki-68101303

2. Registering ramdisk image Execute the following commands to bundle and register the ramdisk image (initrd.img-2.6.28-11-server):
$ euca-bundle-image -i initrd.img-2.6.28-11-server
$ euca-upload-bundle -b mybucket -m /tmp/initrd.img-2.6.28-11-server.manifest.xml
$ euca-register mybucket/initrd.img-2.6.28-11-server.manifest.xml
IMAGE eri-A26613DD

3. Registering disk image Execute the following commands to bundle and register the disk image (centos.5-3.x86.img):
$ euca-bundle-image -i centos.5-3.x86.img --kernel eki-68101303 --ramdisk eri-A26613DD
$ euca-upload-bundle -b mybucket -m /tmp/centos.5-3.x86.img.manifest.xml
$ euca-register mybucket/centos.5-3.x86.img.manifest.xml
IMAGE emi-F62F1100 [5]

4. The new images that have been uploaded can be seen by using following command.
$ euca-describe-images

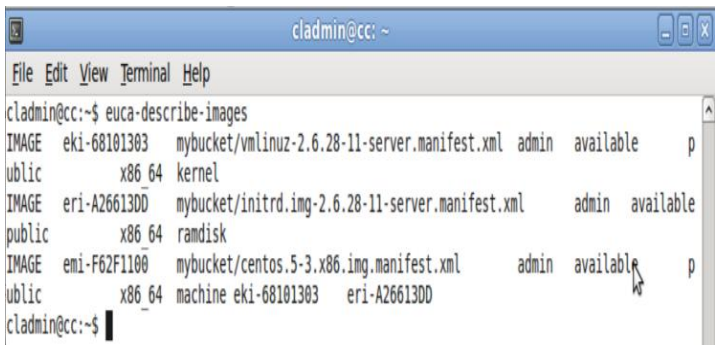
Fig 8a: Screenshot of Images

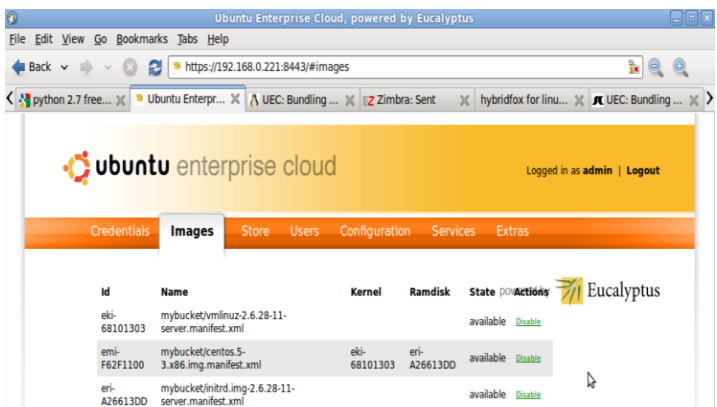
Fig 8b: Screenshot of Images

4.6 Installing a keypair

1. Create a new group called "wiki" and use it instead of the default group
$ euca-add-group wiki -d wiki_demo_appliances

Also allow access to port 22 in instances:
$ euca-authorize wiki -P tcp -p 22 -s 0.0.0.0/0

2. Build a keypair that will be injected into the instance(virtual machine) allowing to access it via ssh.
$ euca-add-keypair mykey > ~/.euca/mykey.priv
$ chmod 0600 ~/.euca/mykey.priv[8]

4.7 Running Instances in cloud

Run instance as follows
$ euca-run-instances -g wiki -k mykey -t c1.medium emi-F62F1100

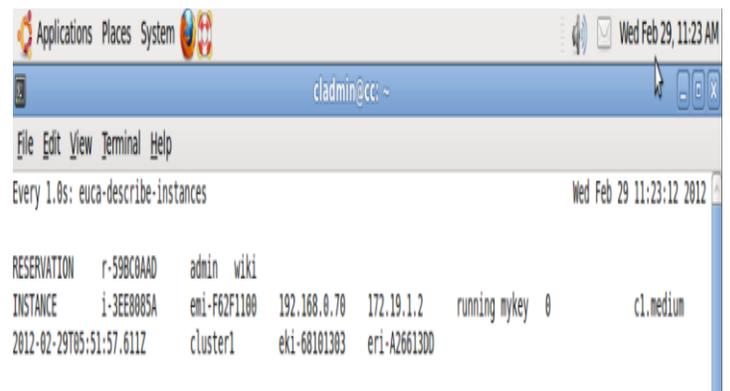
Fig 9: Screenshot of an Instance

4.8 Elasticfox

Elasticfox is an open source Mozilla Firefox extension that works on Firefox Version 2.0 or later to help managing Amazon EC2. This was originally made for EC2 but from version 1.7, Elasticfox added Eucalyptus support as well, because API of Eucalyptus is compatible with that of EC2.
Features of Elasticfox
1. List available Machine images (AMIs in the case of AWS and EMIs in the case of Eucalyptus)
2. List running instances
3. Launch new instances of an AMI/EMI
4. Manage security groups and rules
5. Manage Snapshots/EBS volumes[5]

$ euca-describe-instances

RESERVATION r-518808F3    admin    default
INSTANCE    i-3D7307B3    emi-F62F1100    192.168.0.73    172.19.1.4    running 123 0 m1.large    2012-02-29T10:55:07.374Z cluster1eki-68101303    eri-A26613DD

```
RESERVATION  r-32B7071F        admin    default
INSTANCE     i-45A1094E        emi-F62F1100     192.168.0.71
             172.19.1.3        running  mykey        0        m1.small
             2012-02-29T10:05:10.04Z  cluster1 eki-68101303    eri-A26613DD
```

Now let's observe the above output in an Elasticfox as follows.

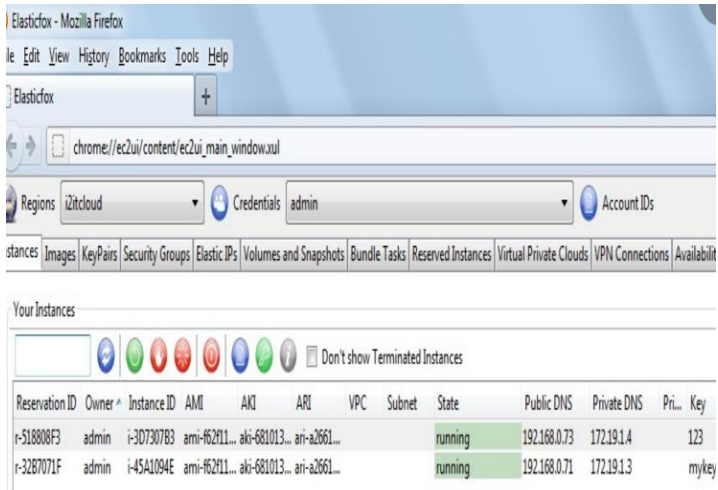

Fig 10: Screenshot of Instances in Elasticfox

### 4.9 Instances

At launch, Instances are in pending state after some time they are in running state at this state client can use the CentOS operating system. An instance to get terminated it first enters into shutting-down state then it terminates[5]

## 5.0 Conclusion

In this way paper describes implementation of a private cloud for an organization providing Infrastructure as a service (IaaS) and providing operating system as Platform as a service (PaaS). Cloud computing will bring flexibility in IT services in coming decade.

**First Author**: He is pursuing **M.Tech** in Advanced Information Technology with specialization in Software Technologies from IGNOU – I[2]IT Centre of Excellence for Advanced Education and Research, Pune, India. He is also **Teaching Assistant** in Advanced Software and Computing Technologies department.He has pursued **B.E** Electronics and Telecommunications from University of Mumbai. His research interests include Cloud computing, Databases and Networking.

**Second Author**: He is pursuing **M.Tech** in Advanced Information Technology with specialization in Software Technologies from IGNOU – I[2]IT Centre of Excellence for Advanced Education and Research, Pune, India. He was **Lab Assistant** in Advanced Software and Computing Technologies department.He has pursued **B.E** Information Technology from University of Nagpur. His research interests include Operating systems, Programming languages and Cloud computing.

**Third Author**: He is pursuing **M.Tech** in Advanced Information Technology with specialization in Software Technologies from IGNOU – I[2]IT Centre of Excellence for Advanced Education and Research, Pune, India. He has pursued **B.E** Information Technology from University of Mumbai. His research interests include databases and cloud computing.

**Fourth Author**: He pursued **Ph.D.** from University of Linz Austria. He is Vice President **Sunflower Information Technologies Pvt. Ltd** Pune, India. He was faculty(Adjunct) in IGNOU – I[2]IT Centre of Excellence for Advanced Education and Research, Pune, India. His research interests include Mathematics, Business Intelligence, Parallel computing and Cloud computing.

**Fifth Author**: He is Senior Executive IT Infrastructure Services in IGNOU – I[2]IT Centre of Excellence for Advanced Education and Research Pune, Maharashtra, India. He pursued **BCA** from GIMT Delhi, India. His key interests are in Virtualization and Network security.